\documentclass[usenatbib,letters]{mnras}
\usepackage{amssymb,amsmath}
\usepackage{graphicx}
\usepackage{xcolor,natbib}
\usepackage{multirow}
\usepackage[T1]{fontenc}
\usepackage{ae,aecompl}
\usepackage{newtxtext, newtxmath}
\usepackage{float}
\usepackage{subfigure}
\usepackage{longtable}
\usepackage{enumitem}
\usepackage{longtable,listings}
\usepackage[flushleft]{threeparttable}
\usepackage{parcolumns}
\def\oiii{[O~{\sc iii}]\ }
\def\nii{[N~{\sc ii}]\ }
\def\oi{[O~{\sc i}]\ }
\def\sii{[S~{\sc ii}]\ }
\def\oiiihb{[O~{\sc iii}]$\lambda5007$\AA$/$H$\beta$\ }

\title[Obscured broad \oiii in Type-2 AGN]
{Evidence for Obscured broad \oiii Components in Type-2 AGN}
\author[Zhang X. G.]
       {Xue-Guang Zhang$^\star$\\
       School of Physics and technology, Nanjing Normal University,
          No. 1, Wenyuan Road, Nanjing, 210023, P. R. China}

\date{}

\begin{document}

\pagerange{\pageref{firstpage}--\pageref{lastpage}} \pubyear{2020}

\maketitle\label{firstpage}

\begin{abstract}  
	In the manuscript, we report evidence on broad \oiii components apparently 
obscured in Type-2 AGN under the framework of the Unified model, after checking properties 
of broad \oiii emissions in large samples of Type-1 and Type-2 AGN in SDSS DR12. We can 
well confirm the statistically lower flux ratios of the broad to the core \oiii components 
in Type-2 AGN than in Type-1 AGN, which can be naturally explained by stronger obscured broad 
\oiii components by central dust torus in Type-2 AGN, unless the Unified model for AGN was 
not appropriate to the narrow emission lines. The results provide further evidence to support 
broad \oiii components coming from emission regions nearer to central BHs, and also indicate 
the core \oiii component as the better indicator for central activities in Type-2 AGN, 
due to few effects of obscuration on the core \oiii component. Considering the broad \oiii 
components as signs of central outflows, the results provide evidence for strong central 
outflows being preferentially obscured in Type-2 AGN. Furthermore, the obscured broad \oiii 
component can be applied to explain the different flux ratios of \oiiihb between Type-1 and 
Type-2 AGN in the BPT diagram.
\end{abstract}


\section{Introduction}

      Type-1 AGN (broad line Active Galactic Nuclei) and Type-2 AGN (narrow line AGN) having 
different observational phenomena can be well explained by orientation effects of central dust 
torus, in the framework of the well-known constantly being revised Unified Model \citep{an93, 
nh15, aa17}. Central broad line regions (BLRs) with tens to hundreds of light-days \citep{kas00, 
ben13, fa17} to central black holes (BHs) are totally obscured by central dust torus in Type-2 
AGN. However, narrow emission line regions (NLRs) with hundreds to thousands of pcs (parsecs) 
to central BHs \citep{fi13, ha14, sg17} lead to expected similar properties of narrow emission 
lines in both Type-1 and Type-2 AGN. Therefore, properties of narrow emission lines can be well 
applied to estimate central activities in Type-2 AGN, such as the reported strong linear 
correlation between AGN power-law continuum luminosity and \oiii luminosity \citep{za03, hb14}.

    Recently, \citet{zh17} have reported the broad \oiii emission regions nearer to central 
BLRs, based on the tighter correlation between AGN continuum luminosity and luminosity of broad 
\oiii components, through a larger sample of SDSS (Sloan Digital Sky Survey) blue quasars (Type-1 
AGN). Actually, broad \oiii components in AGN have been studied for more than three decades. 
\citet{gh05} have shown further effects of central BH potential on broad \oiii components. Similar 
blue broad \oiii components can also be found in \citet{km08, sr11, bl18, so18}. More recently, 
\citet{dh18} have shown interesting results on broader and more blue-shifted broad \oiii emissions 
in the obscured AGN indicating more powerful AGN-driven outflows, a probable challenge to the 
Unified Model of AGN. Furthermore, besides kinematic study on broad \oiii components, geometric 
properties of broad \oiii components have been also well studied in the literature. \citet{sg17} 
have shown that the extended narrow emission regions related to broad \oiii components have 
typically smaller sizes than the sizes of normal \oiii emission regions expected by AGN luminosity 
and/or \oiii luminosity \citep{lz13, ha13, ha14}. Besides plenty of research results on properties 
of broad \oiii emissions in AGN in the literature, we here will focus on one another interesting 
point on broad \oiii emissions in AGN.

    If broad \oiii components were nearer to central BHs, broad \oiii components could be more 
likely to be obscured by central dust torus in Type-2 AGN under the framework of the Unified model.  
The results could provide further information on linkage between narrow and broad emission lines, 
and furthermore, could provide further clues on AGN selection criterion through applications of 
narrow emission line ratios in BPT diagrams \citep{bpt81, ka03, ke06, ke13, kn19, zh20} with line 
ratios on \oiii lines. Meanwhile, further considerations should be given on applications of \oiii 
properties to trace central AGN activities in Type-2 AGN. The manuscript is organized as 
follows. In Section 2, we show our main data samples of both Type-1 and Type-2 AGN. In Section 
3, we show our main results and necessary discussions. In Section 4, we give our final conclusions. 
And in the manuscript, we have adopted the cosmological parameters of 
$H_{0}=70{\rm km\cdot s}^{-1}{\rm Mpc}^{-1}$, $\Omega_{\Lambda}=0.7$ and $\Omega_{\rm m}=0.3$.

\section{Main Samples of Type-2 AGN and Type-1 AGN}

\begin{figure*}
\centering\includegraphics[width = 18cm,height=3.85cm]{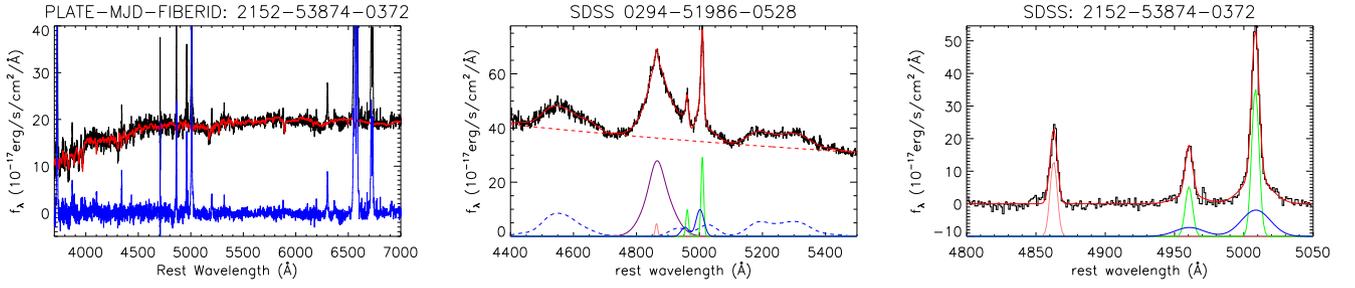}
\caption{Left panel shows an example on the SSP method determined stellar components in the 
Type-2 AGN SDSS 2152-53874-0372. Solid lines in black, red and blue show the observed spectrum, 
the best determined stellar components and the pure line spectrum, respectively. Middle and right 
panels show two examples on the best fitted results to the emission lines around H$\beta$ in the 
Type-1 AGN SDSS 0294-51986-0528 and in the Type-2 AGN SDSS 2152-53874-0372. In middle and right 
panels, solid lines in black and red show the line spectrum and the best fitted results, solid 
pink line shows the determined narrow H$\beta$, solid lines in green and blue show the determined 
core and broad \oiii components, respectively. In middle panel, dashed red line shows the 
determined power law continuum emissions, solid purple line shows the determined broad H$\beta$ 
component, dashed blue line shows the determined optical Fe~{\sc ii} emissions, respectively. 
}
\label{ssp}
\end{figure*}

\begin{figure}
\centering\includegraphics[width = 7cm,height=6cm]{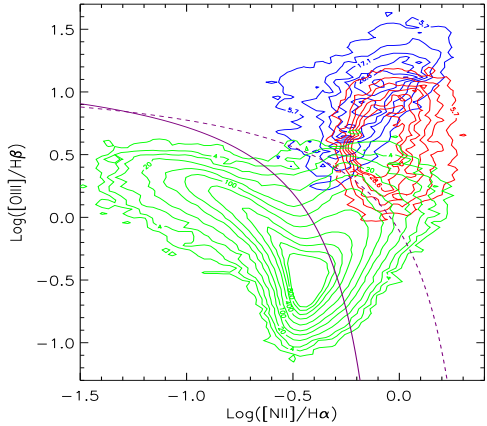}
\caption{BPT diagram for the 5437 Type-1 AGN (contour in blue) and for the 6587 Type-2 AGN (contour 
in red) with reliable narrow emission lines. Solid and dot-dashed purple lines represent the dividing 
lines discussed in \citet{ka03, ke06}. Contour in green shows the results based on more than 240000 
narrow-emission-line galaxies in SDSS DR12.
}
\label{bpt}
\end{figure}

    In the manuscript, Type-1 AGN are collected from the SDSS pipeline classified 
quasars. Type-2 AGN are collected from the SDSS sub-classified AGN in SDSS pipeline classified 
main galaxies based on the MPA-JHU measurements (\url{http://www.sdss3.org/dr9/algorithms/galaxy_mpa_jhu.php}). 
Accepted the criterion of redshift less than 0.3, there are 12342 Type-1 AGN from SDSS quasars and 
16269 Type-2 AGN from SDSS main galaxies collected from SDSS DR12. Then, emission line parameters 
are measured as follows. 
   
    For Type-2 AGN and a small number of Type-1 AGN, of which spectra have clear host galaxy 
contaminations, the widely accepted SSP (Simple Stellar Population) method has been firstly 
applied to subtract the stellar lights, in order to find more accurate emission line properties. 
Detailed descriptions of the SSP method can be found in \citet{bc03, ka03, cm05} etc.. The same 
procedure has been applied in our previous studies in \citet{zh14, zh16, ra17, zh19}, etc.. We 
do not discuss the SSP method any more, but the left panel of Fig.~\ref{ssp} shows an example on 
the SSP method determined stellar components in the Type-2 AGN with PLATE-MJD-FIBERID = 2152-53874-0372.

    After subtractions of the stellar lights, line parameters can be well measured. The emission 
lines are mainly considered around H$\beta$ (rest wavelength from 4400\AA\ to 5600\AA) and around 
H$\alpha$ (rest wavelength from 6250\AA\ to 6850\AA), which are fitted simultaneously by the following 
model functions through the Levenberg-Marquardt least-squares minimization technique (the MPFIT 
package), similar as what we have done in \citet{zh16, zh17}. There are two (or more if necessary, 
after checking the fitted results) broad Gaussian functions applied to describe each broad Balmer 
line (especially in Type-1 AGN), one narrow Gaussian function applied to each narrow emission line 
including narrow Balmer lines, \oiii, \nii, \oi and \sii doublets, and two additional Gaussian 
components applied to describe probably broad components of \oiii doublet (broad \oiii components), 
one broad Gaussian function applied to describe weak He~{\sc ii} line, two power law functions 
applied to describe probable AGN continuum emissions underneath the broad H$\beta$ and underneath 
the broad H$\alpha$, and the Fe~{\sc ii} template discussed in \citet{kp10} applied to describe 
optical Fe~{\sc ii} lines (especially for Type-1 AGN). When the model functions are applied, the 
following restrictions have been accepted, (1) narrow emission lines have the same redshift, (2) 
corresponding broad components in broad Balmer lines have the same redshift, (3) flux ratio of the 
\oiii (\nii) doublet is fixed to the theoretical value 3, (4) there are the same line widths of 
narrow Balmer lines (\oiii or \oi or \nii or \sii doublets), but different line widths for 
different narrow lines. Middle and right panels of Fig.~\ref{ssp} show two examples on the best 
fitted results to the emission lines around H$\beta$. 

    Based on the measured parameters, two criteria have been accepted to collect Type-2 AGN with 
reliable narrow emission lines but no broad Balmer lines. First, measured stellar velocity 
dispersions and line parameters of the narrow emission lines ([O~{\sc iii}]$\lambda5007$\AA (at 
least core \oiii components), narrow Balmer lines and [N~{\sc ii}]$\lambda6583$\AA) are at least 
5 times larger than their corresponding uncertainties. Second, measured line fluxes of broad 
Balmer lines are less than 5 times of the corresponding uncertainties. Then, there are 6587 
Type-2 AGN collected. Moreover, two criteria have been accepted to collect Type-1 AGN with 
reliable narrow and broad emission lines. First, the measured continuum luminosity and line 
parameters of broad Balmer components are at least 5 times larger than their corresponding 
uncertainties. Second, the measured line parameters of [O~{\sc iii}]$\lambda5007$\AA (at least 
core \oiii components), narrow Balmer lines and [N~{\sc ii}]$\lambda6583$\AA\ are at least 5 
times larger than their corresponding uncertainties. Then, there are 5437 Type-1 AGN collected. 
Fig.~\ref{bpt} shows the BPT diagram of flux ratio of [O~{\sc iii}]$\lambda5007$\AA\ to narrow 
H$\beta$ (O3HB) versus flux ratio of [N~{\sc ii}]$\lambda6583$\AA\ to narrow H$\alpha$ (N2HA) 
for the collected 5437 Type-1 AGN and 6587 Type-2 AGN. Here, the [O~{\sc iii}]$\lambda5007$\AA\ 
flux means the total \oiii line flux. The collected objects can be safely classified as AGN, 
based on the dividing lines well discussed in \citet{ka03, ke06, ke13}.

\begin{figure}
\centering\includegraphics[width = 8cm,height=6cm]{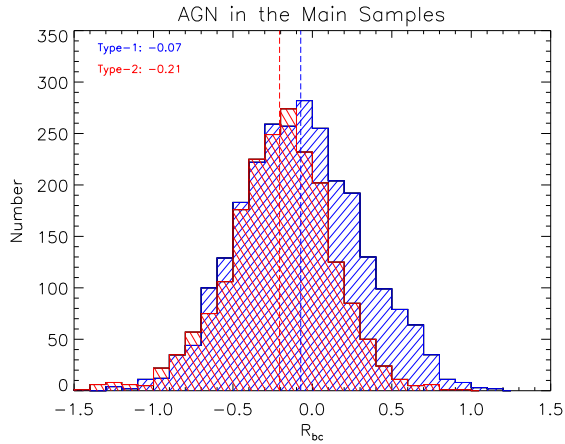}
\caption{Distributions of $R_{\rm bc}$ of the Type-1 AGN (in blue) and the Type-2 AGN (in red) 
in the main samples. Vertical dashed lines in blue and in red show the mean value positions of 
the Type-1 and Type-2 AGN, respectively. Mean value of each distribution is marked in the 
top-left corner.}
\label{ct}
\end{figure}

\begin{figure*}
\centering\includegraphics[width = 18cm,height=8cm]{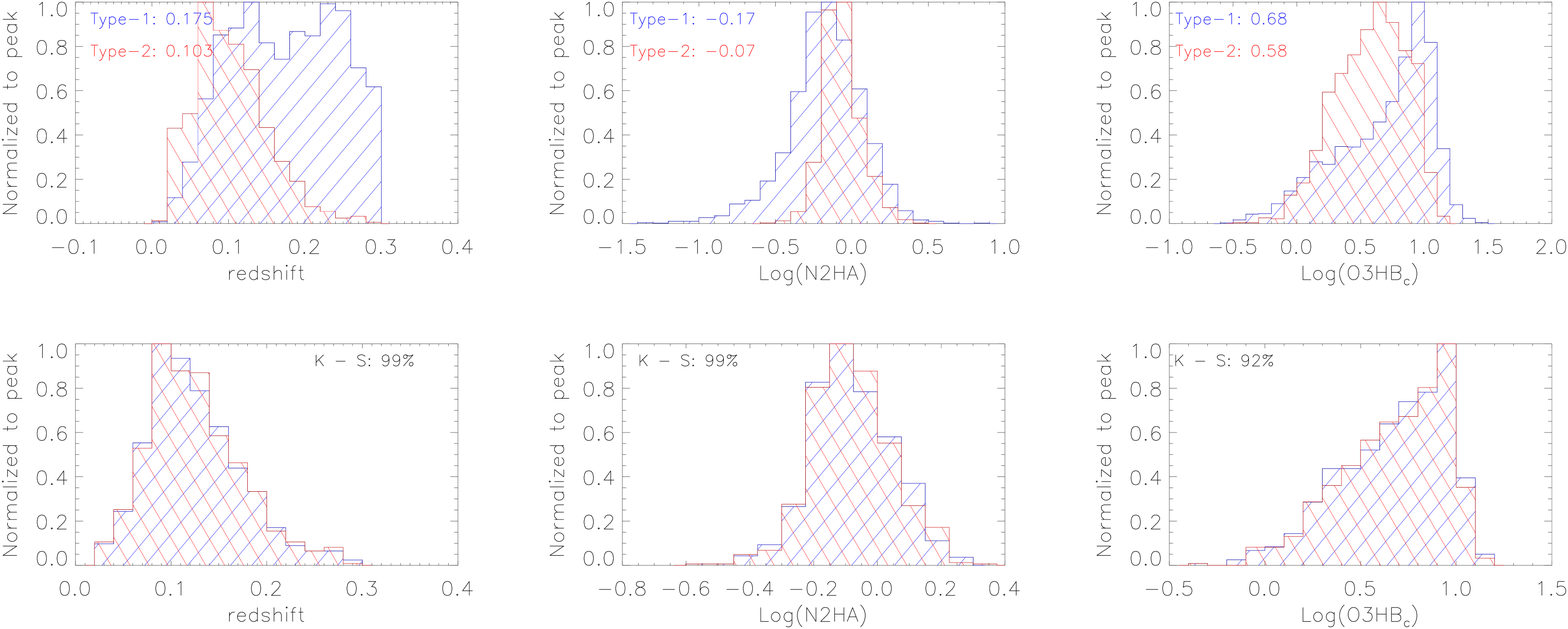}
\caption{Distributions of redshift, N2HA and ${\rm O3HB_c}$ of the Type-1 AGN (in blue) and the 
Type-2 AGN (in red) in the main samples (top panels) and in the BPT/redshift-matched samples 
(bottom panels). In each top panel, mean values of the distributions are marked in the top-left 
corner. In each bottom panel, the calculated Kolmogorov-Smirnov statistical significance level 
is marked in the top-left corner.}
\label{dis_par}
\end{figure*}

\begin{figure}
\centering\includegraphics[width = 8cm,height=10cm]{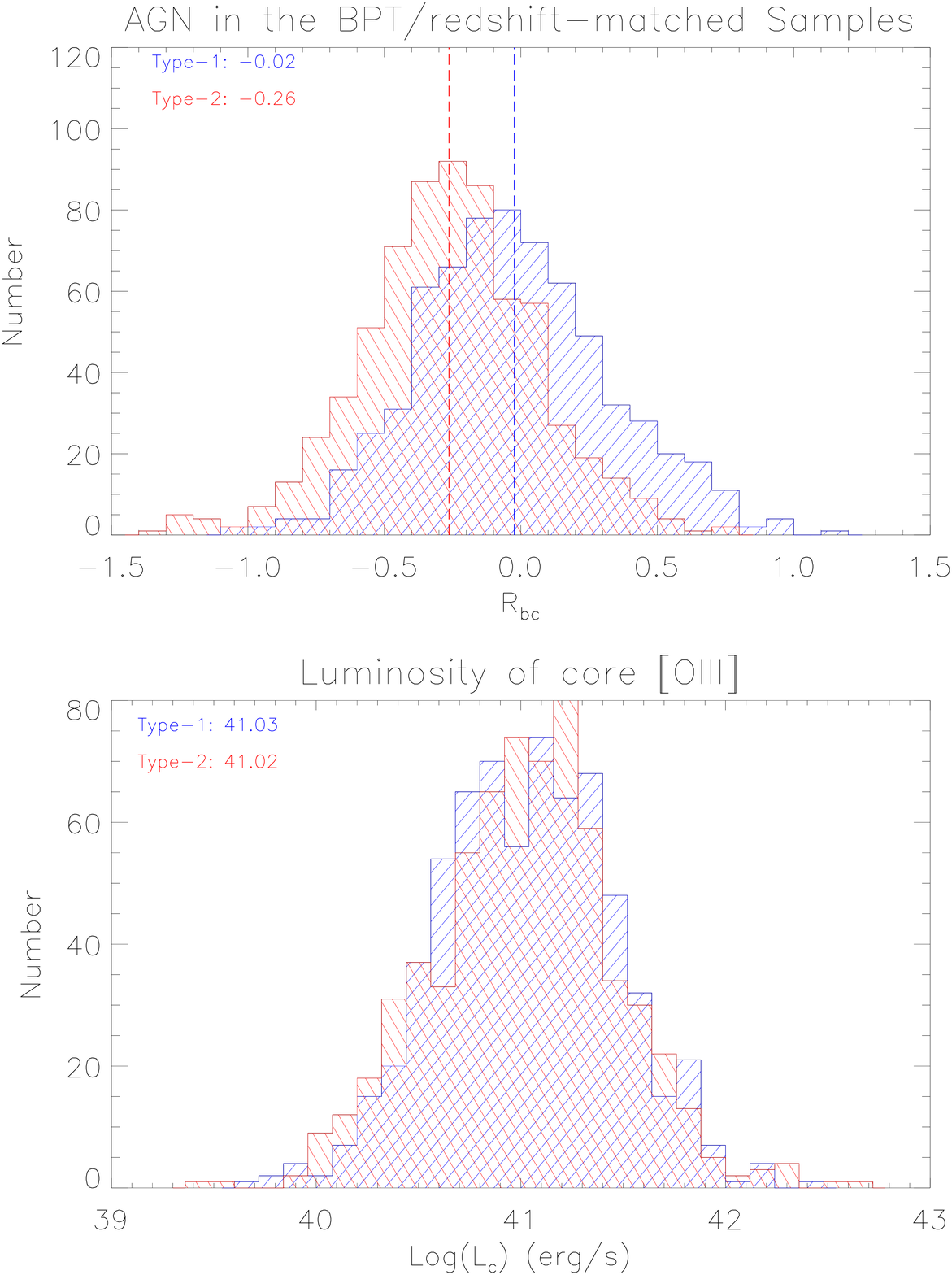}
\caption{Top panel shows the results similar as those in Fig.~\ref{ct}, but for the AGN in 
the BPT/redshift-matched samples. Bottom panel shows luminosity distributions of $\log(L_{\rm c})$ 
of the AGN in the BPT/redshift-matched samples.}
\label{ct_match}
\end{figure}

     Finally, among the 5437 Type-1 AGN and the 6587 Type-2 AGN, the following criteria are 
applied to create our main samples of Type-1 and Type-2 AGN with reliable broad \oiii components: 
the measured line flux and line width (second moment) of both the core and the broad \oiii 
components are at least 5 times larger than their corresponding uncertainties. Here, the determined 
\oiii component with larger second moment is the broad \oiii component. Then, in final main samples, 
there are 2621 Type-1 AGN and 1987 Type-2 AGN, with reliable broad \oiii components.

\section{Main Results and Discussions}

    Based on the determined core and broad \oiii components, properties of the parameter 
$R_{\rm bc}=\log(L_{\rm b})-\log(L_{\rm c})$ have been checked and shown in Fig.~\ref{ct}, 
where $L_{\rm b}$ and $L_{\rm c}$ represent luminosities of the broad and the core \oiii 
components, respectively. The mean $R_{\rm bc}$ are about -0.07 and -0.21\ in the Type-1 and 
Type-2 AGN, respectively. And the Student's T-statistic technique shows that the different 
mean values are significant with levels higher than $10\sigma$. Therefore, there could be 
intrinsic different properties of \oiii components between Type-1 and Type-2 AGN.

      In order to show more accurate results with contaminations as less as possible, The 
following effects have been mainly considered. As the results shown in the top panels of 
Fig.~\ref{dis_par}, the Type-1 and Type-2 AGN have much different distributions of redshift, 
O3HB and N2HA. The different redshift distribution will lead to much different luminosity 
properties of \oiii components. And, the different distributions of O3HB and N2HA will indicate 
much different central activities between the Type-1 and Type-2 AGN in the main samples. In 
order to totally ignore effects of different distributions of redshift and emission line 
ratios between the Type-1 and Type-2 AGN, a simple method has been considered, by comparing 
Type-1 and Type-2 AGN in two subsamples which have the same distributions of redshift and 
emission line ratios (called BPT/redshift-matched samples).

     Based on the distributions of redshift, ${\rm O3HB_c}$ (flux ratio of the core \oiii 
component to narrow H$\beta$) and N2HA of the 2621 Type-1 AGN and the 1987 Type-2 AGN in 
the main samples, 667 Type-1 AGN and 667 Type-2 AGN are randomly collected from the main 
samples to create the two BPT/redshift-matched samples which have the same distributions of 
redshift, ${\rm O3HB_c}$ and N2HA, as the results shown in the bottom panels of 
Fig.~\ref{dis_par}. Here, not O3HB but ${\rm O3HB_c}$ is applied, because of the broad 
\oiii components in Type-2 AGN probably obscured. Based on the strong correlation between 
core \oiii luminosity and AGN continuum luminosity in \citep{zh17}, central activities 
can also be well traced by applications of core \oiii components. The two-sided 
Kolmogorov-Smirnov statistic technique has been applied to confirm the same distributions 
of redshift, ${\rm O3HB_c}$ and N2HA with significance levels higher than 92\% between 
the 667 Type-1 AGN and the 667 Type-2 AGN in the BPT/redshift-matched samples.

    Then, the parameter $R_{\rm bc}$ has been re-checked in the top panel of Fig.~\ref{ct_match} 
for the AGN in the BPT/redshift-matched samples, with the mean $R_{\rm bc}$ of about -0.02 and 
-0.26\ in the 667 Type-1 AGN and in the 667 Type-2 AGN, respectively. The difference is more 
apparent than the results for the AGN in the main samples shown in Fig.~\ref{ct}. And the different 
mean values have Student's T-statistic determined significance levels higher than $10\sigma$. 
Furthermore, the bottom panel of Fig.~\ref{ct_match} shows the luminosity distribution of 
$L_{\rm c}$ for the AGN in the BPT/redshift-matched samples with the same mean values. And through 
the two-sided Kolmogorov-Smirnov statistic technique, the Type-1 and Type-2 AGN in the 
BPT/redshift-matched samples have the same $L_{\rm c}$ distributions with significance levels higher 
than 60\%. Therefore, intrinsic different broad \oiii components lead to the different $R_{\rm bc}$ 
between the Type-1 and Type-2 AGN, and the direct and natural explanation to the lower $L_{\rm b}$ in 
Type-2 AGN is that the broad \oiii emissions have been obscured, under the framework of the Unified 
model for AGN.

    Before proceeding further, there is one point we should note. In the manuscript, we do 
not discuss the effects of beam smearing on our results. The beam smearing effects have been 
discussed for more than five decades \citep{bk89, wl09, gr10, ss16, zw17}, especially on kinematic 
properties through integral-filed spectra. The more recent discussions on the effects of beam 
smearing can be found in \citet{hh16, hh20}. For the SDSS optical fiber spectra discussed in the 
manuscript, it is hard to clearly determine and remove the beam smearing effects on emission 
lines, due to loss of physical information of spatially resolved velocity field. In order to 
roughly check the beam smearing effects, line width (second moment) difference 
$\Delta_{bc}=\sigma_b-\sigma_c$ between the broad and core \oiii components can be roughly 
applied to show properties of central velocity gradient. Then, the dependence of $\Delta_{bc}$ 
on the parameter $R_{\rm bc}$ have been checked in the Type-1 and Type-2 AGN in the 
BPT/redshift-matched samples. The Spearman rank correlation coefficients are about -0.12 and 
0.04 for the Type-1 and Type-2 AGN, respectively, strongly indicating no dependence of the 
parameter of $R_{\rm bc}$ on central velocity gradient in Type-1 AGN nor in Type-2 AGN. Therefore, 
the beam smearing effects have few effects on our final results, even there are different 
beam smearing effects between Type-1 and Type-2 AGN due to different orientation effects.

    Meanwhile, we provide further discussions on the obscured broad \oiii components in Type-2 
AGN. First, similar as results discussed in \citet{zh17}, the core rather than the broad \oiii 
components (or the total \oiii) could be the better indicator to central activities in Type-2 
AGN. Through the parameter of $R_{\rm bc}$ different in Type-1 and Type-2 AGN, we can roughly 
estimate that about 50\% of broad \oiii emissions are obscured in Type-2 AGN. Therefore, the 
classification by narrow emission line ratios in the BPT diagram should lead to lower O3HB, if 
total \oiii lines were considered. Second, based on similar intrinsic properties of \oiii emission 
components expected by the framework of the Unified model for AGN, there are few selection effects 
on the results shown in Fig.~\ref{ct_match}. Third, based on the results shown in Fig.~\ref{bpt}, 
the Type-2 and Type-1 AGN have statistical different O3HB in the BPT diagram, with mean 
$\log({\rm O3HB})$ about 0.91 and 0.71\ in the Type-1 and Type-2 AGN, respectively. Once, we 
simply accepted that about 50\% of broad \oiii emissions are obscured in Type-2 AGN by properties 
of $R_{\rm bc}$, we could expect the intrinsic flux ratio of $\log({\rm O3HB})$ in Type-2 AGN 
about $0.71+\log(1.5)\sim0.89$ similar as the mean value of 0.91 of the Type-1 AGN.

     As the direct and natural explanation on weaker broad \oiii emissions in Type-2 AGN by 
obscuration, it will be interesting to consider sources of the obscurations. Central dust 
torus in AGN could be preferred, rather than randomly moving dust clouds in central regions, 
otherwise, there should be similar obscurations on broad \oiii emissions in Type-1 AGN. As 
discussed properties of central dust torus in \citet{bo15, gh15, zh18}, opening angle could 
be around $\theta\sim40-60$\degr\ in AGN, and the dust sublimation radius 
$R_{\rm sub}\propto1.3{\rm pc}\times L^{0.5}$ 
could be accepted as the radius of dust torus. In order to provide appropriate obscurations on 
broad \oiii emissions in Type-2 AGN, scales of the distance $R_{\rm B3}$ of broad \oiii emission 
regions to central BHs could be simply around $R_{\rm B3}\sim R_{\rm sub}\times\tan(\theta/2)$. 
If a global mean value of $\theta\sim25$\degr was accepted, we will have 
$R_{\rm B3}\sim0.6{\rm pc}(L_{\rm UV}/10^{46}{\rm erg/s})^{0.5}$. 
Meanwhile, the BLRs size can be well estimated as 
$R_{\rm BLRs}\propto34(L_{\rm opt}/10^{44}{\rm erg/s})^{0.5}{\rm light-days}$ \citep{kas00, ben13}. 
Then, an oversimplified result can be expected $R_{\rm B3}\sim20\times R_{\rm BLRs}$, about 3 
magnitudes smaller than the distance of common \oiii emission regions to central BHs \citep{ha13}, 
providing interesting clues on very broader \oiii components in AGN. Certainly, we should 
note that the expression $R_{\rm B3}\sim20\times R_{\rm BLRs}$ is estimated
through tremendously oversimplified structures of central dust torus, however, the results
can provide structure information on the potential obscured broad \oiii emission regions
much nearer to central BHs between BLRs and normal NLRs in AGN, as expected by properties
of kinematically-disturbed broad \oiii regions related to outflows discussed in \citet{sg17}.

   Before the end of the section, line intensity properties rather than kinematic properties 
of broad and core \oiii components are mainly considered in the manuscript. More detailed discussions 
can be found on kinematic properties of gas outflows through properties of \oiii emissions in 
Type-2 AGN in \citet{woo16, woo17} and in Type-1 AGN in \citet{dh18}. Here, we simply check the 
correlation between stellar velocity dispersions ($\sigma_\star$) and line widths of broad 
($\sigma_b$) and core \oiii ($\sigma_c$) components in the Type-2 AGN, and find that the mean 
ratios of $\sigma_\star$ to $\sigma_c$ and of $\sigma_\star$ to $\sigma_b$ are about 1.05 and 0.27, 
respectively. The results well consistent with the reported results in \citet{woo16} strongly 
indicate broad \oiii components tightly related to outflowing gases in Type-2 AGN. And the wider 
broad \oiii components can be well expected due to the broad \oiii emission regions with deeper 
gravitational potential nearer to central BHs. Moreover, we check the expected positive correlation 
between line width of the broad \oiii components and the \oiii luminosity in Type-1 and Type-2 AGN, 
such as the positive correlations shown in Figure~6\ in \citet{dh18}. Here, the broad \oiii luminosity 
is applied to trace the central AGN luminosity as discussed in \citet{zh17} in Type-1 AGN, but 
the core \oiii luminosity is applied in Type-2 AGN. Then, positive correlations can be found with 
Spearman rank correlation coefficients about 0.40 with $P_{\rm null}\sim10^{-23}$ and about 
0.32 with $P_{\rm null}\sim10^{-17}$ in the Type-1 AGN and in the Type-2 AGN, respectively. It 
is clear that the kinematic properties of the collected Type-1 and Type-2 AGN are consistent 
with the reported results in the literature. 

\section{conclusions}   

     Finally, we give our main conclusions as follows. Based on the SDSS high quality spectra 
of large samples of Type-1 and Type-2 AGN with reliable broad \oiii components, different 
properties of broad \oiii components can be confirmed between the Type-1 and Type-2 AGN: 
statistically lower broad \oiii luminosities and statistically lower $R_{\rm bc}$ (flux ratio 
of the broad to the core \oiii component) in the Type-2 AGN, after considering necessary effects. 
The results indicate stronger obscuration on the broad \oiii components in the Type-2 AGN due to 
broad \oiii emission regions nearer to central BHs, under the framework of the Unified model for 
AGN. Considering the broad \oiii components as robust signs of central outflows, the results 
provide evidence for obscured central outflows in Type-2 AGN. Moreover, rather than total \oiii 
lines, the core \oiii components can be treated as the better indicator of central activities in 
Type-2 AGN, due to few effects of obscuration on the core \oiii components. Furthermore, the 
obscured broad \oiii components can be well applied to explain the different flux ratios of 
O3HB in the BPT diagram between Type-1 and Type-2 AGN.

\section*{Acknowledgements}
Zhang gratefully acknowledges the anonymous referee for giving us constructive 
comments and suggestions greatly improving our paper. Zhang gratefully thanks the kind 
financial support from Nanjing Normal University and the kind grant support from NSFC-11973029. 
This paper has made use of the data from the SDSS projects. The SDSS-III web site is 
http://www.sdss3.org/. SDSS-III is managed by the Astrophysical Research Consortium for the 
Participating Institutions of the SDSS-III Collaboration.

\section*{Data Availability}
The data underlying this article will be shared on reasonable request to the corresponding 
author (\href{mailto:xgzhang@njnu.edu.cn}{xgzhang@njnu.edu.cn}).

\bsp
\label{lastpage}
\end{document}